%% file: ICRC2025_template_IceCube.tex
\documentclass[a4paper,11pt]{article}
\usepackage{float}
\usepackage{pos}
\usepackage{lipsum} 
\usepackage{lineno}
\usepackage{titlesec}
\titlespacing*{\section}{3pt}{3pt}{3pt}
\titlespacing*{\subsection}{2pt}{2pt}{2pt}
\titlespacing*{\subsubsection}{2pt}{2pt}{2pt}
\usepackage{wrapfig}
\usepackage{graphicx}

\title{Characterizing the Astrophysical Neutrino Flux Using Contained and Uncontained Cascade Events in IceCube}

\ShortTitle{Characterizing the Astrophysical Neutrino Flux Using Contained and Uncontained Cascade Events}

\author{The IceCube Collaboration \\{\normalsize \normalfont(a complete list of authors can be found at the end of the proceedings)}\\}

\emailAdd{zrechav@icecube.wisc.edu}
\emailAdd{emre.yildizci@icecube.wisc.edu}
\emailAdd{lu.lu@icecube.wisc.edu}

\abstract{
Recently, the IceCube Neutrino Observatory has reported a deviation from the single power law in the extragalactic diffuse neutrino flux. A neural network-based event selection of contained and uncontained cascade events from IceCube, in which uncontained events have interaction vertices at the edge or outside of the detector instrumentation volume, has a factor \textasciitilde3 gain in effective area over the cascade events used in the novel combined tracks and cascades selection which reported the deviation. Systematic improvements and rigorously updated modeling of the atmospheric neutrino background is incorporated into this high statistics contained and uncontained cascade event selection to clarify features of the astrophysical neutrino spectrum across energies from 1 TeV up to 100 PeV.

\vspace{4mm}

{\bfseries Corresponding authors:}
Zoë Rechav$^{1*}$, 
Emre Yildizci$^{1}$, 
Lu Lu$^{1}$ 
\\
{$^{1}$ \itshape University of Wisconsin-Madison}\\
[4mm]
$^*$ Presenter
}

\input{ICRCdetails.tex}

\begin{document}

\maketitle
\section{Introduction}
The IceCube Neutrino Observatory has discovered and characterized the astrophysical neutrino flux. The Enhanced Starting Track Event Selection (ESTES) \cite{PhysRevD.110.022001} finds a single power law $\Phi(E) = \Phi_{0} \times \left(\frac{E}{E_0}\right)^{-\gamma}$ (SPL) to be the best fit, while the Medium Energy Starting Events (MESE) \cite{MESE} and Combined Fit \cite{Naab:2023v2} selections find broken power law (BPL), deviating from the SPL below \textasciitilde30 TeV. The BPL model is shown in Eq.~\ref{equations}, where $\Phi_{b}$ is the flux normalization assuming a fixed break energy (E$_{b}$), and $\gamma_{1,2}$ the spectral indices for BPL below and above E$_{b}$, respectively:
\begingroup
\setlength{\abovedisplayskip}{3pt}
\setlength{\belowdisplayskip}{3pt}
\begin{equation}
\Phi(E) = \Phi_{b} \times 
\begin{cases}
{
\begin{aligned}
&\left(\frac{E}{E_b}\right)^{-\gamma_1} & E < E_b \\
&\left(\frac{E}{E_b}\right)^{-\gamma_2} & E \geq E_b
\end{aligned}
\text{(BPL) .
}}
\end{cases}
\label{equations}
\end{equation}
\endgroup
\noindent
Further characterization of the astrophysical neutrino flux using an increased statistics sample is needed to resolve features of the spectrum. 

While the ESTES, MESE, and Combined Fit selections select reconstructed events contained within or passing through the instrumented detector volume \cite{Aartsen:2016nxy}, a cascade-like selection of both contained and uncontained events built using neural networks and used to discover neutrino flux emission from the galactic plane \cite{Sclafani:2022}, known as Deep Neural Network Cascades (DNNCascades), could be the solution. IceCube has performed previous analyses using uncontained events above PeV energies \cite{Glashow}, but DNNCascades' method of removing through-going muon background allows for the inclusion of uncontained events down to 1 TeV and increases statistics of contained events below 10 TeV.

This work describes the optimization of DNNCascades to characterize the diffuse astrophysical spectrum down to 1 TeV energy. Updated systematic treatment of the Antarctic ice \cite{ftpv3}\cite{leo:2025icrc} and atmospheric flux \cite{Ya_ez_2023}, improved energy reconstruction \cite{leo:2025icrc}, and new quality cuts are employed into the selection. Additionally, enhanced parametrization of the atmospheric neutrino flux accepted by IceCube predicts a reduced atmospheric neutrino background in the southern sky, improving the astrophysical purity of the selection. The selection, known as DNNCascades Diffuse, is presented here as an increased statistics selection of contained and uncontained events that could resolve features of the diffuse astrophysical neutrino spectrum down to 1 TeV.
\section{Event Selection}
The sample consists of cascade-like events with reconstructed energies as low as 1 TeV, selected through a series of deep neural network algorithms. Two machine learning-based classifiers were used to finalize event morphologies of the selection. Each classifier predicts a score per event to reject atmospheric muon background (Muonness) and identify cascade-like events (Cascadeness), respectively \cite{Sclafani:2022}. The resulting selection consists of cascade-like, neutral current muon neutrinos, and electron and tau neutrinos. 

Quality cuts based on detector geometry were employed to preserve angular and energy resolution of the selection. Removed events are in regions  where limited knowledge of ice and atmospheric flux impact event reconstruction. The cuts include removing events with reconstructed vertices above the instrumented detector volume, inside the dust layer \cite{Aartsen:2016nxy}, events below the instrumented detector volume with energy < 10 TeV, and any event greater than 700 meters radially from the detector center. Detailed event rates of the astrophysical and atmospheric neutrino flux after applying the classifiers and quality cuts are shown in Table \ref{cuts}. Muon background event rates are subdominant in the selection and not displayed here.
\begin{table}[H]
\centering
\resizebox{\linewidth}{!}{
\begin{tabular}{|c|c|c|c|}
\hline
Quality Cut & Specification & $\nu_{astro}$ event rate [Hz] & $\nu_{conv}$ event rate [Hz] \\ \hline
Selection + & & 3.275$\times10^{-5}$ & 3.076$\times10^{-4}$  \\ \hline
Muonness & score < 0.005 & 3.073$\times10^{-5}$ & 2.825$\times10^{-4}$  \\ \hline
Cascadeness & score > 0.1 & 2.436$\times10^{-5}$ & 1.151$\times10^{-4}$ \\ \hline
Surface & z$_{reco}$ < 500m & 2.430$\times10^{-5}$ & 1.150$\times10^{-4}$ \\ \hline
Dust Layer & z$_{reco}$ !$\epsilon$[50,150]m & 2.361$\times10^{-5}$ & 1.132$\times10^{-4}$  \\ \hline
Deep Ice & E$_{reco}$ > 10 TeV if z$_{reco}$ $\epsilon$[-500,-700]m & 2.150$\times10^{-5}$ & 9.748$\times10^{-5}$ \\ \hline
\end{tabular}
}
\caption{Event rate of topology and quality cuts for DNNCascades Diffuse, after applying each cut in succession to existing cuts from DNNCascades (Selection +). The expected astrophysical and conventional neutrino event rates assume the SPL model of Cascades with $\gamma = 2.53$ and $\Phi_{0} = 1.66$ \cite{PhysRevLett.125.121104} and the GaisserH4a atmospheric flux model \cite{Gaisser_2012}. For reference, the IceCube Neutrino Observatory is roughly 1 cubic km with a detector center defined in cartesian coordinates (0,0,0) \cite{Aartsen:2016nxy}.}
\label{cuts}
\end{table}
\subsection{Contained and Uncontained Events}
\begin{figure}[H]
    \centering
    \includegraphics[width=.8\linewidth]{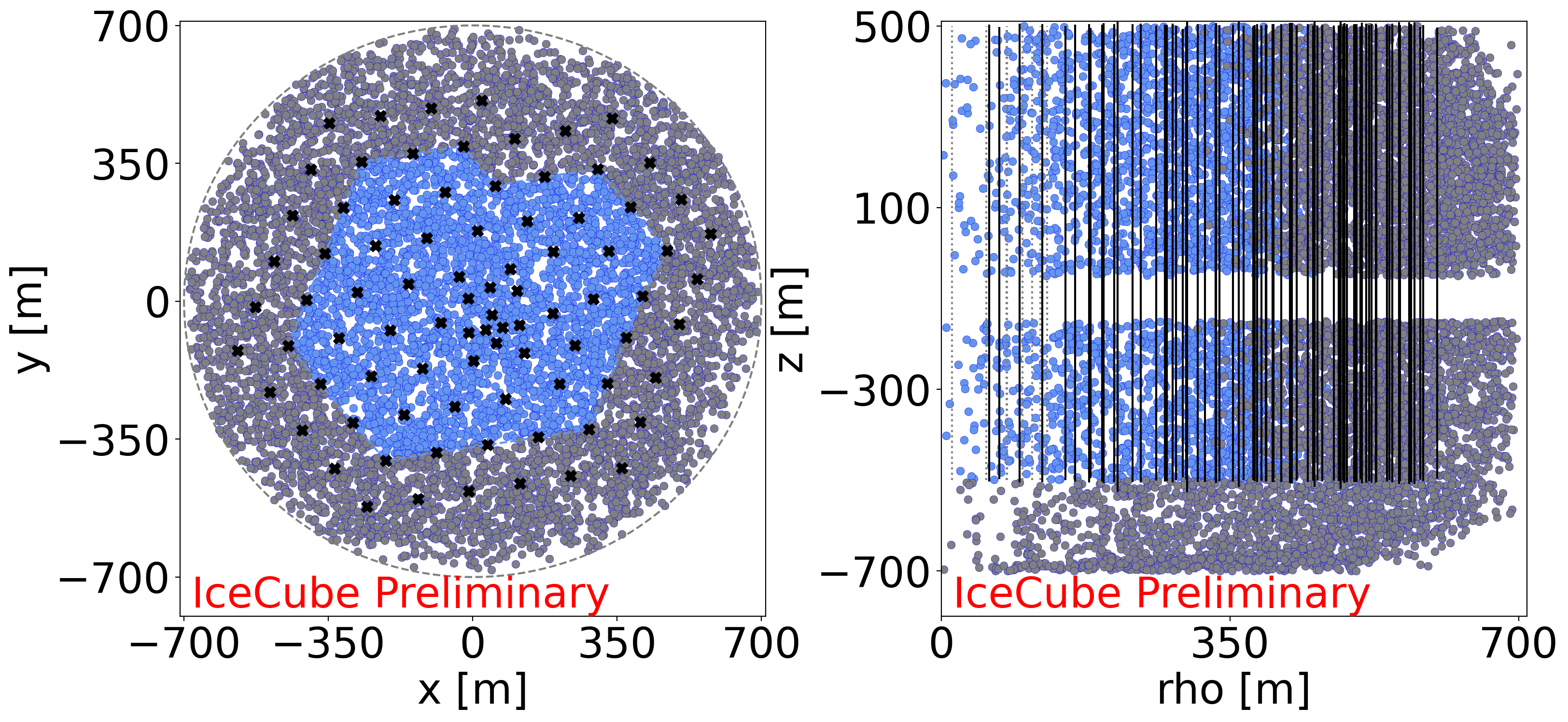}
    \caption{Defining the contained and uncontained region for IceCube events in cartesian and cylindrical coordinates for DNNCascades Diffuse selection. Electron neutrinos with primary energy 1 - 100 PeV are displayed, with contained events in blue and uncontained events in gray. IceCube strings are displayed in black, and DeepCore strings as dotted lines in the right panel \cite{Aartsen:2016nxy}. The relationship between depth [m] and z [m] is depth = 1948 - z, where 1948 is the distance in meters from the detector centor to the ice surface \cite{Aartsen:2016nxy}.}
    \label{geometry}
\end{figure}

Events are characterized as contained or uncontained in DNNCascades Diffuse to determine statistical gain and reconstruction quality in uncontained events . Displayed in Figure~\ref{geometry}, a contained event is defined as an event with reconstructed vertex inside the instrumented volume of the IceCube detector. This includes all events within 1.4km to 2.5km depth, and within the second outermost layer of digital optical module (DOM) \cite{Aartsen:2016nxy} strings. An Uncontained event is defined in Figure~\ref{geometry} as any event outside of the contained region and extends down to 2.7km depth and within a reconstructed radial distance from the detector center to 700m. Assuming the SPL model from Cascades with $\gamma = 2.53$ and $\Phi_{norm} = 1.66$ \cite{PhysRevLett.125.121104}, DNNCascades Diffuse MC is made of \textasciitilde60\% contained events and \textasciitilde40\% uncontained events. 

\begin{figure}[H]
    \centering
    \includegraphics[width=.9\linewidth]{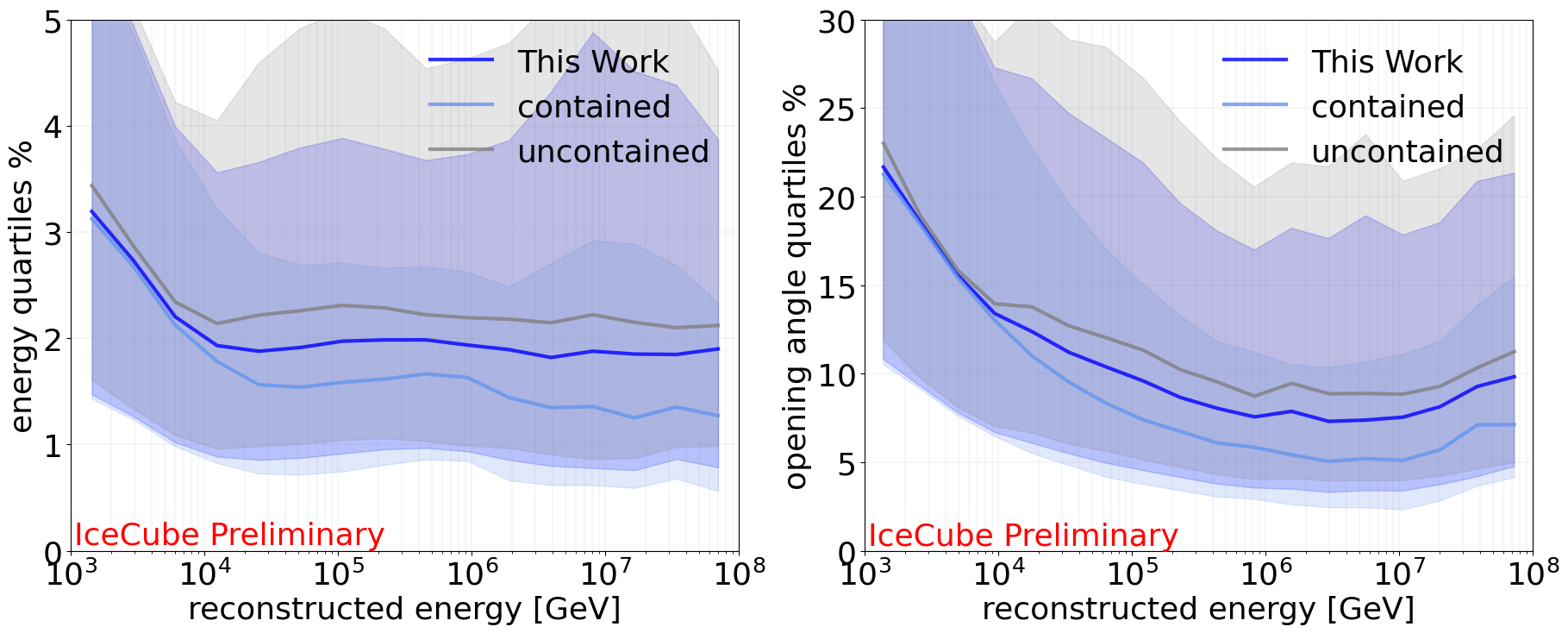}
    \caption{Angular and energy resolutions for DNNCascades Diffuse, total, contained, and uncontained events.}
    \label{resolution}
\end{figure}

Reconstruction quality is displayed in Figure~\ref{resolution}. Uncontained events are expectedly reconstructed with poorer resolution than contained events. When the reconstructed vertex is farther from the detector, fewer photons are detected by DOMs, which can lead to overestimation of reconstructed energy due to uncertainties in reconstructed position. No distinct decrease in radial energy resolution was observed except at distances greater than 150m from the instrumented detector volume.
\subsection{Effective Area}
\begin{figure}[H]
  \centering
  \begin{minipage}{0.50\linewidth}
    \includegraphics[width=\linewidth]{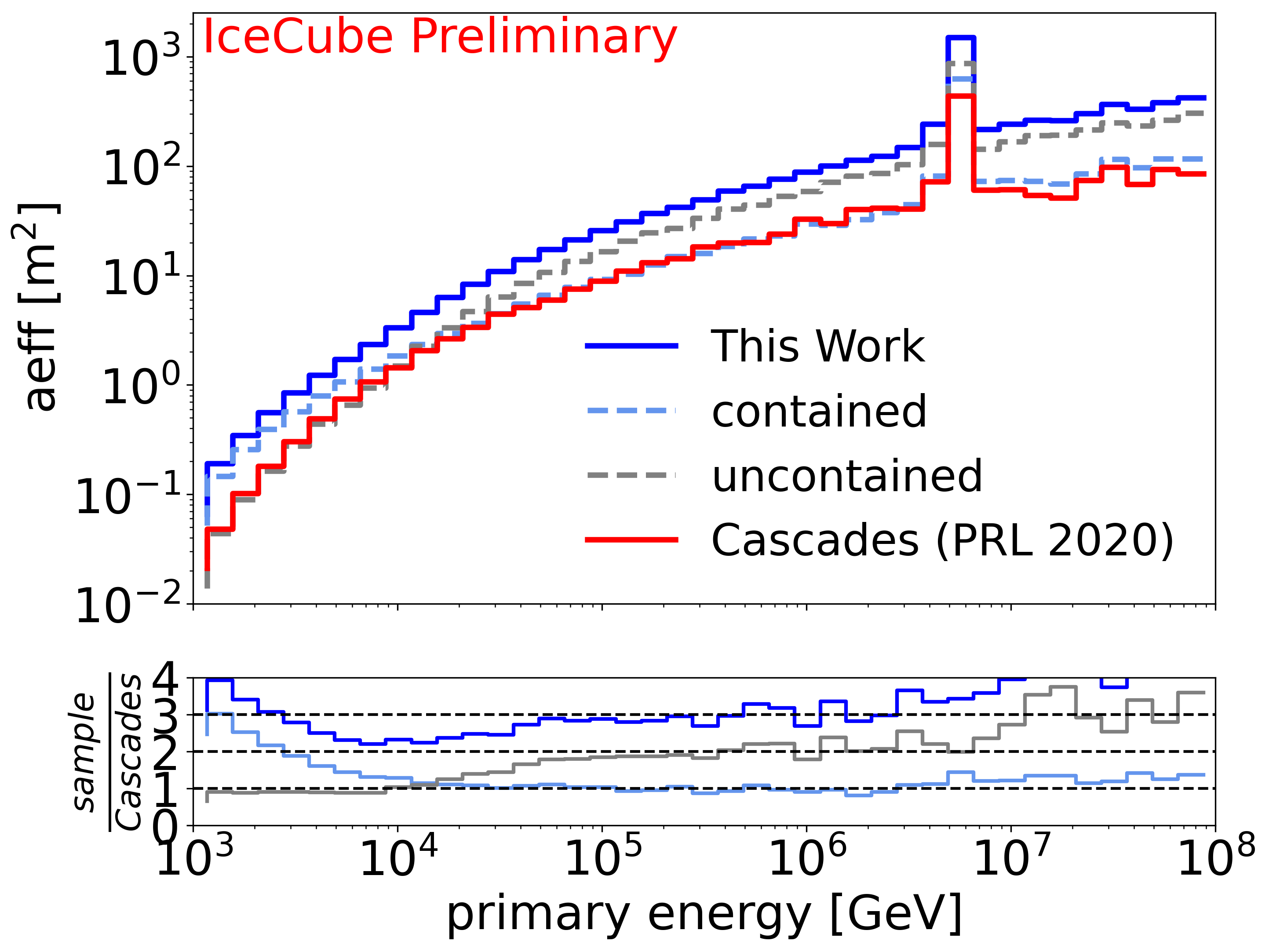}
  \end{minipage}
  \hspace{0.05\linewidth}
  \begin{minipage}{0.4\linewidth}
    \caption{Effective area of DNNCascades Diffuse, contained and uncontained cascades, and Cascades \cite{PhysRevLett.125.121104}. The increase in effective area for DNNCascades Diffuse, total, contained, and uncontained cascades samples, are compared to Cascades.}
    \label{aeff}
  \end{minipage}
\end{figure}

The increase in sensitivity of DNNCascades Diffuse can be visualized by comparing the effective area of the sample to the contained-only cascades sample, Cascades \cite{PhysRevLett.125.121104}, used in the Combined Fit \cite{PhysRevLett.125.121104}. Figure~\ref{aeff} shows contained and uncontained events of DNNCascades Diffuse compared to Cascades. The overall gain in effective area for DNNCascades Diffuse is between a factor \textasciitilde2 and \textasciitilde4 depending on the observed energy range.
\section{Modeling of the Atmospheric Neutrino Flux}
At energies below \textasciitilde30 TeV, modeling the dominant atmospheric flux enhances differentiation from the astrophysical flux. In particular, Cosmic Ray interactions produce showers containing atmospheric neutrinos and muons that reach the IceCube detector in coincidence. Not all atmospheric neutrinos reach IceCube with detectable muons and simulating air showers with CORSIKA \cite{CORSIKA} at IceCube's TeV sensitivity is computationally prohibitive, leading to particle-separate simulation. The phenomenon of atmospheric neutrino rejection due to coincident muons, known as the self veto \cite{Gaisser_2014}, is not observed in separated simulation. IceCube models this self veto effect via the passing fraction (P\textsubscript{\it{pass}}), which estimates the probability that an atmospheric neutrino survives event selection with undetectable accompanying muons \cite{Gaisser_2014}\cite{Arg_elles_2018}.

The open source self veto parametrization package NuVeto \cite{Arg_elles_2018} is used to calculate P\textsubscript{\it{pass}} as a function of neutrino energy, zenith angle, depth, and primary particle shower development, described in detail in \cite{Gaisser_2014}\cite{Arg_elles_2018}, is displayed here Eq.~\ref{pass_probability}. An important parameter of the P\textsubscript{\it{pass}} is the detector response to accompanying muons (P\textsubscript{\it{light}}). In practice, P\textsubscript{\it{light}} is the fraction of muons that trigger the rejection of a neutrino event:
\begingroup
\setlength{\abovedisplayskip}{3pt}
\setlength{\belowdisplayskip}{3pt}
 {\fontsize{10}{12}\selectfont
\begin{align}
\it{P_{pass}}\left(E_{\nu},\theta_{z}\right) &= \frac{1}{\Phi_\nu\left(E_{\nu}, \theta_{z}\right)} \sum_{A}\sum_{\it{p}}\int dE_{\it{p}} \int \frac{dX}{\lambda_{\it{p}}\left(E_{\it{p}},X\right)}\int dE_{CR}\frac{dN_{A,\it{p}}}{dE_{\it{p}}}\left(E_{CR},E_{\it{p}},X \right) \Phi_{A}\left(E_{CR}\right) \notag \\
&\quad \times \left[ 1 - \int dE_{\mu}^{i}\frac{dN_{\it{p},\mu}}{dE_{\mu}^{i}}\left(E_{\it{p}},E_{\nu},E_{\mu}^{i}\right)\int dE_{\mu}^{f} \, \int dE_{\mu}^{f} \, P_{\text{reach}}(E_{\mu}^{f} \mid E_{\mu}^{i},\theta_z) \, \underline{P_{\text{light}}(E_{\mu}^{f})}(E_{\mu}^{i}, \theta_z)\right] \notag\\& e^{-N_{A,\mu(E_{CR}-E_{\it{p},\theta_{z}})}}. \label{pass_probability}
\end{align}
 }
 \endgroup
\noindent
In past analyses, P\textsubscript{\it{light}} was approximated using analytical step functions \cite{PhysRevD.110.022001} \cite{PhysRevLett.125.121104} \cite{Arg_elles_2018} or single muon showers\cite{HESE_2021}. In this work, neutrino correlated muon bundles are injected into surviving neutrino events of DNNCascades Diffuse to more closely emulate the full simulation of cosmic showers as seen in nature. The total number of events with muon bundles (N$_{\nu  +\mu, total}$) is compared to the number of events with muon bundles that have passed DNNCascades Diffuse (N$_{\nu + \mu, pass}$). The fractional rate of muon bundles that trigger event rejection is thus given by P\textsubscript{\it{light}} = 1 - $\frac{N_{\nu+\mu, pass}}{N_{\nu + \mu, total}}$. This method of calculating P\textsubscript{\it{light}} was first implemented in MESE \cite{MESE}. Here the method has been expanded and incorporated into DNNCascades Diffuse.
\subsection{Correlated Muon Bundles Using CORSIKA}
A novel method was developed to determine which muon bundles are most likely to accompany atmospheric neutrinos. The multiplicity and energy distribution of the muon bundles used to build $N_{\nu + \mu, total}$ were correlated with a same shower neutrino using CORSIKA simulation \cite{CORSIKA}. 
\subsubsection{Simulation}
 CORISKA airshower simulation generated for this work assumed the Gaisser H3A, 5-component atmospheric density model (Proton, He, N, Al, Fe) \cite{Gaisser_2012} and the Sibyll2.3d hadronic interaction model including charm production \cite{CORSIKA}. The updated simulation accounted for uncontained events in DNNCascades Diffuse, in Figure \ref{geometry} with detectable events up to 700 radial meters from the detector center. From Earth's surface, PROPOSAL \cite{PROPOSAL} was used to propagate particles to the IceCube detector, considering particle interactions and decays, and energy losses. 

 Information from \textasciitilde450 million simulated air showers was collected. The energy, direction ($\cos\theta$), depth, and flavor of one random neutrino per air shower was recorded. Leading muons by energy were collected with a ceiling multiplicity of N=4 due to computational and statistical limitations. No requirement of correlated (sibling) or uncorrelated (family) muon to the sampled neutrino was imposed. 
 \subsubsection{Correlations}
\begin{figure}[H]
    \centering
    \includegraphics[width=1.0\linewidth]{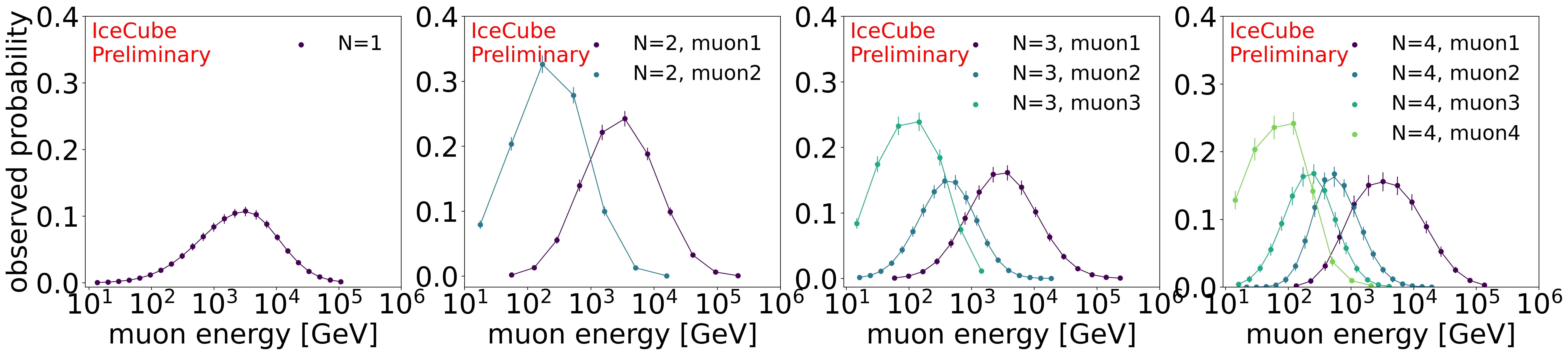}
    \caption{Muon energy distributions for a 160 TeV muon neutrino in the bin $\cos\theta \in [0.6, 0.8]$, at depths $\in [1.6, 2.0]$ km, for muon multiplicities N=1, 2, 3, and 4.}
    \label{corr}
 \end{figure}

The recorded information was binned to determine the most likely muon energies per bundle for a given neutrino by sampled flavor, energy, $\cos\theta$, depth, and muon multiplicity N. In total, there are 720 muon energy distributions. For each distribution, the best fit double-sided log gaussian function was determined using $\chi^{2}$ values. To account for varying statistics in each of the 720 distributions, the number of muon energy bins in each distribution was allowed to float as a fit parameter. An example of the recorded muon energies and their best fits for ten bins are shown in Figure~\ref{corr}, for a 160 TeV muon neutrino in the region $\cos\theta \in [0.6, 0.8]$, and depth $\in [1.6, 2.0]$ km for muon multiplicities N=1, 2, 3, and 4. These distributions correlate the most likely muon bundle multiplicity and energy distribution for a given atmospheric neutrino. 
 \noindent
 \subsection{Detector Response to Muon Bundles}
\begin{figure}[H]
  \centering
  \begin{minipage}{0.45\linewidth}
    \includegraphics[width=\linewidth]{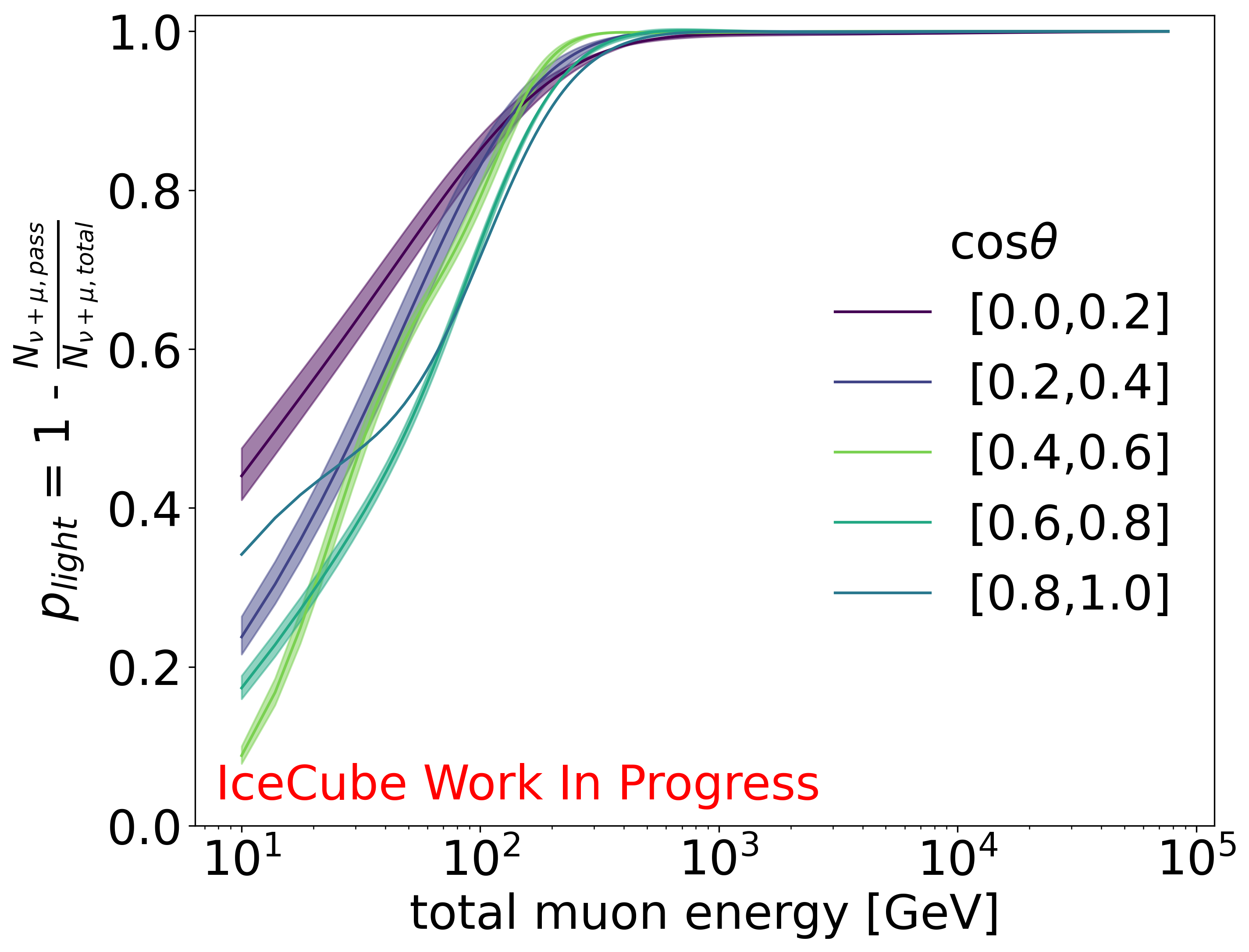}
  \end{minipage}
  \hspace{0.05\linewidth}
  \begin{minipage}{0.45\linewidth}
    \caption{P\textsubscript{\it{light}} distributions separated by cosine zenith bin, over all depths in the detector for muon neutrinos. Uncertainty bands are given by $\pm\sigma$ of the fit parameters a and x1, where a is the sigmoid transition and x1 is the gaussian peak of the modified sigmoid function used to fit each P\textsubscript{\it{light}} distribution. The most downgoing bin has small uncertainty bands.}
    \label{plight}
  \end{minipage}
\end{figure}

 To calculate P\textsubscript{\it{light}}, the number of surviving DNNCascades Diffuse MC neutrinos is given by N$_{\nu}$. For each event in N$_{\nu}$ , muon bundles are sampled from distributions as seen in Figure~\ref{corr} and injected to create N$_{\nu +\mu, total}$. The DNNCascades Diffuse selection is reapplied on N$_{\nu +\mu, total}$, resulting in the surviving number of neutrinos with muon bundles, ${N_{\nu + \mu, pass}}$. The fraction of muon bundles that trigger event rejection, P\textsubscript{\it{light}} = 1 - $\frac{N_{\nu+\mu, pass}}{N_{\nu + \mu, total}}$, is calculated as a function of neutrino region (depth, cos$\theta$) and flavor. 5 of 60 P\textsubscript{\it{light}} distributions calculated to estimate the detector response to muon bundles in DNNCascades Diffuse are shown in Figure~\ref{plight}.
 \noindent
 \subsection{Passing Fraction}
\begin{figure}[H]
     \centering
     \includegraphics[width=.8\linewidth]{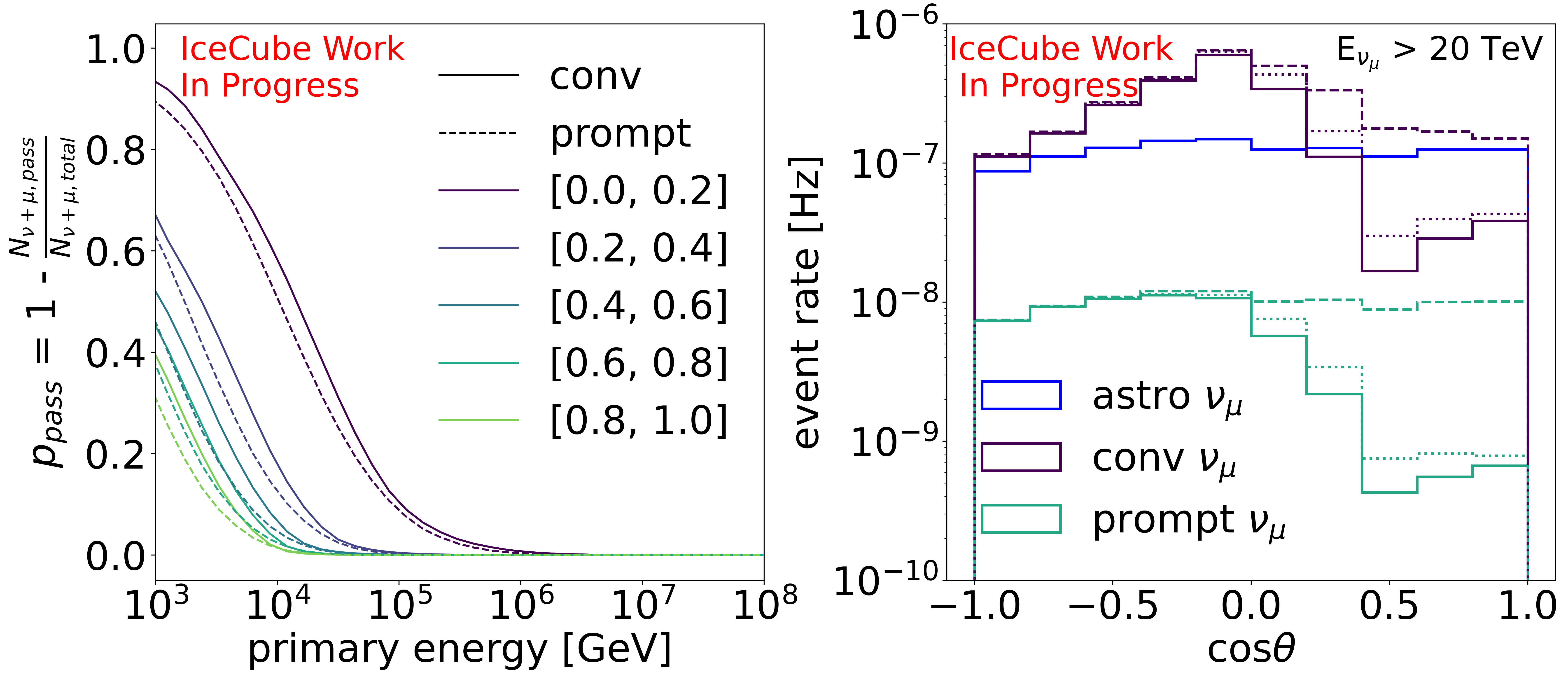}
     \caption{(Left) Passing Fraction as a function of $\cos(\theta)$ for conventional and prompt muon flux.(Right) $\cos(\theta)$ distribution for muon neutrinos in the DNNCascades Diffuse event selection for all muon neutrinos with energy >20 TeV, with updated passing fraction.}
     \label{pass}
 \end{figure}

 The passing fractions for prompt and conventional muon neutrinos including the P\textsubscript{\it{light}} whose updated calculation been described in this work are shown in Figure~\ref{pass} (left) for DNNCascades Diffuse  as a function of $\cos(\theta)$. Horizontal events have longer travel paths to the detector that impacts muon bundle detectability at lower energies, which leads to an increased probability to pass the self veto. At the highest muon energies, correlated to neutrino energies, the angular distribution of arriving muons is isotropic. Additionally, prompt neutrinos are produced via charmed meson decay and are more likely to have higher energy accompanying muons reach the detector due to their differing production mechanisms, resulting in a decreased probability to pass the self veto. 
\noindent
 The impact of the self veto on the atmospheric neutrino flux for muon neutrinos is shown in Figure~\ref{pass} (right). The existing parametrization of the sample is compared with the parametrization described in this work. In the southern sky, implementing the muon bundle into the detector response calculations results in an increased rejection of atmospheric neutrino events and further breakdown in the degeneracy between the prompt and astrophysical flux in the southern sky. After applying the updated passing fraction, the astrophysical flux dominates for energies greater than \textasciitilde20 TeV. 
 \section{Sensitivity}
\begin{figure}[H]
  \centering
  \begin{minipage}{0.55\linewidth}
    \includegraphics[width=\linewidth]{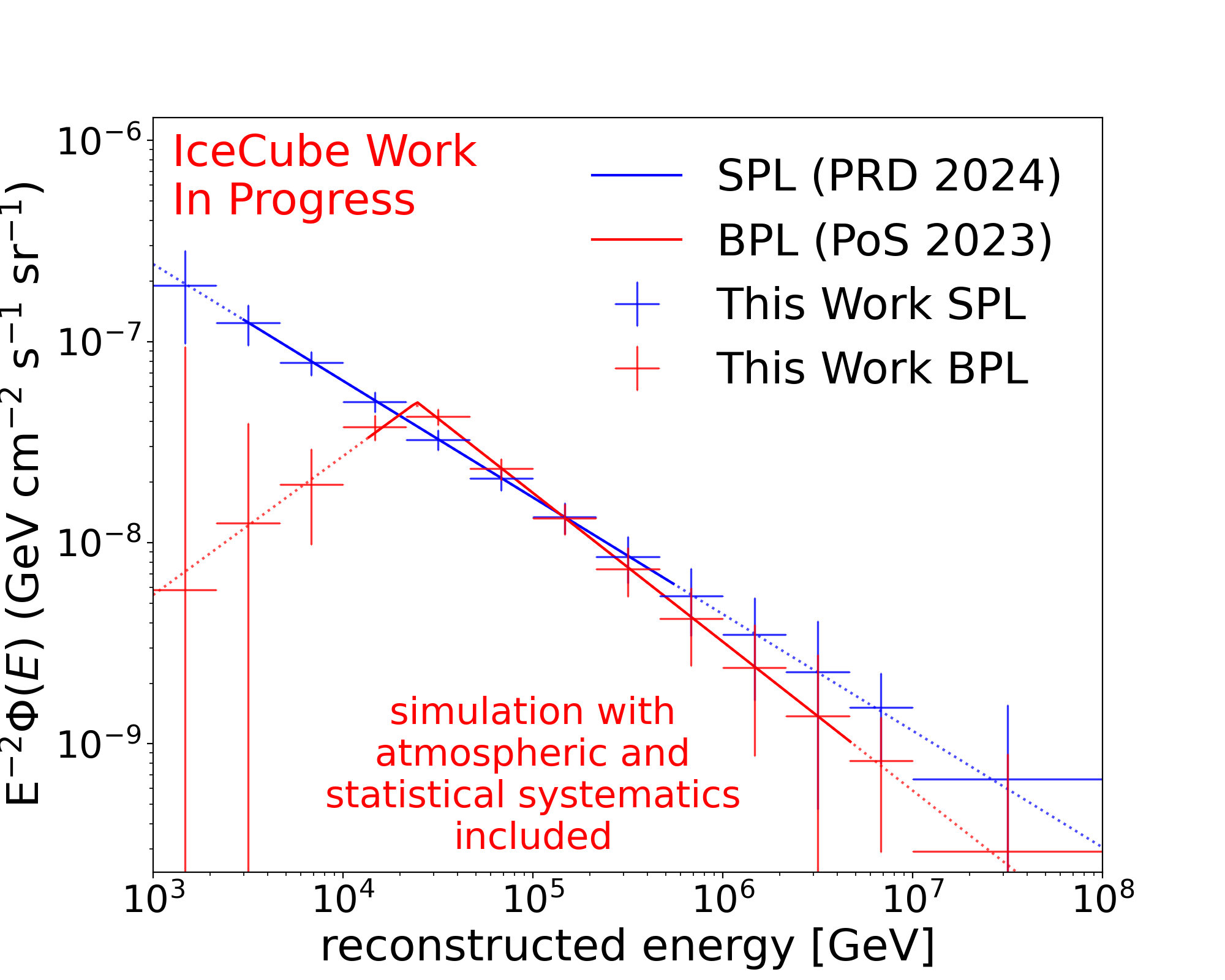}
  \end{minipage}
  \hspace{0.05\linewidth}
  \begin{minipage}{0.35\linewidth}
    \caption{Expected differential fluxes using MC, assuming SPL model from ESTES \cite{PhysRevD.110.022001} and BPL model from Combined Fit \cite{Naab:2023v2}. Sensitive energy range of ESTES [3 TeV, 550 TeV] and Combined Fit [13.7 TeV, 4.7 PeV] shown in solid lines. Atmospheric systematics and statistic uncertainty incorporated, using PoissonLLH method.}
    \label{piecewise}
  \end{minipage}
\end{figure}

 To gauge the sensitivity in differentiating features of the astrophysical spectrum, models of the Combined Fit \cite{Naab:2023v2} and ESTES \cite{PhysRevD.110.022001} were injected to fit differential fluxes for DNNCascades Diffuse. Atmospheric systematics and statistical uncertainties are incorporated into the 1$\sigma$ error bars derived from the Hessian matrix. Figure~\ref{piecewise} shows the results, where the spectral break occurs at approximately 30 TeV. 
 \section{Conclusions}
 DNNCascades Diffuse is a high statistics contained and uncontained event selection with updated treatment of systematic uncertainties of the atmospheric fluxes and the ice, quality cuts based on improved reconstruction resolution, and extensive characterization of the atmospheric neutrino flux in the southern sky. The preliminary results suggest improved sensitivity to resolve features of the astrophysical neutrino spectrum especially below energies of 30 TeV.

\bibliographystyle{ICRC}
\bibliography{references}

\clearpage
\input{authorlist_IceCube.tex}

\end{document}

%% file: ICRCdetails.tex
\ConferenceLogo{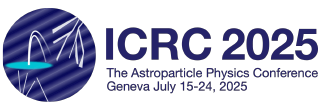}

\FullConference{39th International Cosmic Ray Conference (ICRC2025)\\
 15–24 July 2025\\
Geneva, Switzerland\\}

%% file: authorlist_IceCube.tex
\section*{Full Author List: IceCube Collaboration}

\scriptsize
\noindent
R. Abbasi$^{16}$,
M. Ackermann$^{63}$,
J. Adams$^{17}$,
S. K. Agarwalla$^{39,\: {\rm a}}$,
J. A. Aguilar$^{10}$,
M. Ahlers$^{21}$,
J.M. Alameddine$^{22}$,
S. Ali$^{35}$,
N. M. Amin$^{43}$,
K. Andeen$^{41}$
% ...
\\
\\
% ...
$^{10}$ Universit{\'e} Libre de Bruxelles, Science Faculty CP230, B-1050 Brussels, Belgium \\
% ...
$^{16}$ Department of Physics, Loyola University Chicago, Chicago, IL 60660, USA \\
$^{17}$ Dept. of Physics and Astronomy, University of Canterbury, Private Bag 4800, Christchurch, New Zealand \\
% ...
$^{21}$ Niels Bohr Institute, University of Copenhagen, DK-2100 Copenhagen, Denmark \\
$^{22}$ Dept. of Physics, TU Dortmund University, D-44221 Dortmund, Germany \\
% ...
$^{35}$ Dept. of Physics and Astronomy, University of Kansas, Lawrence, KS 66045, USA \\
% ...
$^{39}$ Dept. of Physics and Wisconsin IceCube Particle Astrophysics Center, University of Wisconsin{\textemdash}Madison, Madison, WI 53706, USA \\
% ...
$^{41}$ Department of Physics, Marquette University, Milwaukee, WI 53201, USA \\
% ...
$^{43}$ Bartol Research Institute and Dept. of Physics and Astronomy, University of Delaware, Newark, DE 19716, USA \\
% ...
$^{63}$ Deutsches Elektronen-Synchrotron DESY, Platanenallee 6, D-15738 Zeuthen, Germany \\
$^{\rm a}$ also at Institute of Physics, Sachivalaya Marg, Sainik School Post, Bhubaneswar 751005, India \\
% ...

\subsection*{Acknowledgments}

\noindent
The authors gratefully acknowledge the support from the following agencies and institutions:
USA {\textendash} U.S. National Science Foundation-Office of Polar Programs,
U.S. National Science Foundation-Physics Division,
U.S. National Science Foundation-EPSCoR,
U.S. National Science Foundation-Office of Advanced Cyberinfrastructure,
Wisconsin Alumni Research Foundation,
Center for High Throughput Computing (CHTC) at the University of Wisconsin{\textendash}Madison,
Open Science Grid (OSG),
Partnership to Advance Throughput Computing (PATh),
Advanced Cyberinfrastructure Coordination Ecosystem: Services {\&} Support (ACCESS),
Frontera and Ranch computing project at the Texas Advanced Computing Center,
U.S. Department of Energy-National Energy Research Scientific Computing Center,
Particle astrophysics research computing center at the University of Maryland,
Institute for Cyber-Enabled Research at Michigan State University,
Astroparticle physics computational facility at Marquette University,
NVIDIA Corporation,
and Google Cloud Platform;
Belgium {\textendash} Funds for Scientific Research (FRS-FNRS and FWO),
FWO Odysseus and Big Science programmes,
and Belgian Federal Science Policy Office (Belspo);
Germany {\textendash} Bundesministerium f{\"u}r Forschung, Technologie und Raumfahrt (BMFTR),
Deutsche Forschungsgemeinschaft (DFG),
Helmholtz Alliance for Astroparticle Physics (HAP),
Initiative and Networking Fund of the Helmholtz Association,
Deutsches Elektronen Synchrotron (DESY),
and High Performance Computing cluster of the RWTH Aachen;
Sweden {\textendash} Swedish Research Council,
Swedish Polar Research Secretariat,
Swedish National Infrastructure for Computing (SNIC),
and Knut and Alice Wallenberg Foundation;
European Union {\textendash} EGI Advanced Computing for research;
Australia {\textendash} Australian Research Council;
Canada {\textendash} Natural Sciences and Engineering Research Council of Canada,
Calcul Qu{\'e}bec, Compute Ontario, Canada Foundation for Innovation, WestGrid, and Digital Research Alliance of Canada;
Denmark {\textendash} Villum Fonden, Carlsberg Foundation, and European Commission;
New Zealand {\textendash} Marsden Fund;
Japan {\textendash} Japan Society for Promotion of Science (JSPS)
and Institute for Global Prominent Research (IGPR) of Chiba University;
Korea {\textendash} National Research Foundation of Korea (NRF);
Switzerland {\textendash} Swiss National Science Foundation (SNSF).